\documentclass[pra,showpacs,superscriptaddress,twocolumn,eqsecnum]{revtex4}

\usepackage{amsfonts,amsmath,amssymb,subfigure,graphicx,bm,times,color}

\newcommand{\suma}[1]{\sum_{{#1} \in \mathbb{Z}}}
\newcommand{\intcirc}{\int_{2\pi}}
\newcommand{\op}[1]{\hat {#1}}
\newcommand{\Tr}{\mathop{\mathrm{Tr}}\nolimits}

\begin{document}

\title{Wigner function for twisted photons}

\author{I.~Rigas}
\affiliation{Departamento de \'{O}ptica, Facultad de F\'{\i}sica,
Universidad Complutense, 28040 Madrid, Spain}

\author{L. L.~S\'{a}nchez Soto}
\affiliation{Departamento de \'{O}ptica, Facultad de F\'{\i}sica,
Universidad Complutense, 28040 Madrid, Spain}

\author{A. B.~Klimov}
\affiliation{Departmento de F\'isica, Universidad de Guadalajara, 
44420 Guadalajara, Jalisco, Mexico}

\author{J.~\v{R}eh\'{a}\v{c}ek}
\affiliation{Department of Optics, Palacky University, 
17. listopadu 50, 772 00 Olomouc, Czech Republic}

\author{Z~Hradil}
\affiliation{Department of Optics, Palacky University, 17. listopadu 50,
772 00 Olomouc, Czech Republic}

\date{\today}

\begin{abstract}
  A comprehensive theory of the Weyl-Wigner formalism for the
  canonical pair angle-angular momentum is presented, with special
  emphasis in the implications of rotational periodicity and
  angular-momentum discreteness.
\end{abstract}


\maketitle

\section{Introduction}

A quantum system has a dynamical symmetry group $G$ if its Hamiltonian
is a function of the generators of $G$. In this case, the Hilbert
space of the system splits into a direct sum invariant subspaces
(carriers of the irreducible representations of $G$) and the
discussion of any physical property can be restricted to one of 
these subspaces~\cite{Barut:1987}.

The existence of such a symmetry also allows for the explicit
construction of a phase space for the system as the coadjoint orbit
associated with an irreducible representation of
$G$~\cite{Kostant:1970,Kirillov:1976} (in fact, it turns out to be a
symplectic manifold). In consequence, to every operator on Hilbert
space we can associate a function on phase space, opening the way to
formally representing quantum mechanics as a statistical theory on
classical phase space. Various aspects of this formalism for basic
quantum systems have been developed by a number of
authors~\cite{Weyl:1928,Wigner:1932,Moyal:1949,Stratonovich:1956,
Agarwal:1970,Berezin:1975,Agarwal:1981,Bertrand:1987,Varilly:1989,
Atakishiyev:1998,Brif:1998,Benedict:1999}.

There are, however, important differences with respect to a classical
description. They come from the noncommuting nature of conjugate
quantities, which precludes their simultaneous precise measurement
and, therefore, imposes a fundamental limit to the accuracy with which
we can determine a point in phase space. As a distinctive consequence
of this, there is no unique rule by which we can associate a classical
phase-space variable to a quantum operator and depending on the
operator ordering, various functions can be defined. For example, the
quantum state (i.e., the density matrix) of the system can be mapped
onto a whole family of functions parametrized by a number $s$; the
values $+ 1$, 0, and $-1$ corresponding to the Husimi $Q$, the Wigner
$W$, and the Glauber-Sudarshan $P$ functions, respectively.  These
phase-space functions are known as quasiprobability distributions, 
as  in quantum mechanics they play a role similar to that of genuine 
probability distributions in classical statistical mechanics
(for reviews, see Refs.~\cite{Balazs:1984,Hillery:1984,Lee:1995,Schroek:1996}).

Apart from the description of the harmonic oscillator (for which $G$
is the Heisenberg-Weyl group and the corresponding phase space is the
plane $\mathbb{R}^2$), this formalism has also been successfully
applied to spin-like systems (or qubits in the modern parlance of
quantum information), for which $G$ is the group SU(2) and the phase
space is the two-dimensional Bloch sphere. However, one can rightly
argue that this Wigner function, although describing a discrete
system, is not defined in a discrete phase space. In fact, the growing
interest in quantum information has fueled the search for discrete
phase-space counterparts of the Wigner function (see
Ref.~\cite{Klimov:2008} for a complete and up-to-date review). The
main advantage of such a representation consists in that even states
from different irreducible representations can be pictured on the same
phase space, which is basically a direct product of two-dimensional
discrete tori.

There is still another ``mixed'' canonical pair: angle and angular
momentum.  Now, the symmetry group $G$ is noncompact and can be taken
as the two-dimensional Euclidean group E(2), whereas the associated
phase space is the discrete cylinder $\mathbb{Z} \times
\mathcal{S}_1$ ($\mathcal{S}_1$ denotes here the unit circle), since
one of the variables is continuous and the other is discrete. Several
interesting properties of a number of systems, such as molecular
rotations, electron wave packets, Hall fluids, and light fields, to
cite only a few examples, can be described in terms of this symmetry
group~\cite{Rigas:2008}.  In quantum optics, it is the basic tool to
deal with the orbital angular momentum of the so-called twisted
photons~\cite{Molina:2007,Franke:2008}, which have been proposed 
for applications in quantum experiments~\cite{Vaziri:2002}.
 
The construction of a proper Wigner function for this case is still 
under discussion. Although some interesting attempts have been
published~\cite{Mukunda:1979,Bizarro:1994,Mukunda:2005}, they seem 
of difficult application to practical problems. Quite interesting
group-theoretical approaches to this problem can be also found in
Refs.~\cite{Nieto:1998,Plebanski:2000}. In this paper, we approach 
this interesting problem from the perspective of finite-dimensional 
systems and construct a \textit{bona fide} Wigner function that 
fulfills all the reasonable requirements and is easy to handle and 
to interpret. We also discuss its applications to some relevant
quantum states.

\section{Wigner function for position-momentum}
\label{sec:qpWig}

In this section we briefly recall the relevant structures needed to
set up the Wigner function for Cartesian quantum mechanics. This is 
to facilitate comparison with the angular case later on. For
simplicity, we choose one degree of freedom only, so the associated
phase space is the plane $\mathbb{R}^2$.

The canonical Heisenberg commutation relations between Hermitian
coordinate and momentum operators $\op{q}$ and $\op{p}$ are (in units
$\hbar = 1$)
\begin{equation}
  \label{eq:HWcom}
  [\op{q}, \op{p}] = i \, ,
\end{equation}
so that they are the generators of the Heisenberg-Weyl algebra. In the
unitary Weyl form this is expressed as
\begin{equation}
  \label{eq:Weyl}
  \op{U} (p) \op{V}(q) = \op{V} (q) \op{U}(p) \, e^{i q p} \, ,
\end{equation}
where
\begin{equation}
  \op{V} (q) = \exp(- i q \op{p} ) \, ,
  \qquad
  \op{U} (p) = \exp(i p \op{q} ) \, , 
\end{equation}
are the generators of translations in position and momentum, 
respectively. In the Cartesian case, these exponentials can be
entangled to define a displacement operator
\begin{equation}
  \label{eq:HWDisp1}
  \op{D} (q,p) = \op{U} (p)  \op{V}(q)  e^{- i q p/2} = 
  \exp[i(p \op{q} -  q \op{p})] \, ,
\end{equation}
with the parameters $(q,p)$ labelling phase-space points. However,
this cannot be done for other canonical pairs, as we shall see.  

The displacement operators form a complete trace-orthonormal set (in  
the continuum sense) in the space of operators acting on $\mathcal{H}$ 
(the Hilbert space of square integrable functions on $\mathbb{R}$):
\begin{equation}
  \label{eq:HWDispOrtho}
  \Tr [ \op{D} (q, p) \, \op{D}^\dagger (q^\prime, p^\prime) ] = 
  2 \pi \delta (q - q^\prime) \delta (p - p^\prime)  \,  .
\end{equation}
Note that $\op{D}^\dagger (q, p) = \op{D} (-q, - p)$, while
$\op{D}(0,0) = \op{\openone}$.

The mapping of the density matrix $\op{\varrho}$ into a Wigner
function defined on $\mathbb{R}^2$ is established in a canonical way:
\begin{eqnarray}
  \label{eq:Wigcan}
  & W(q, p)  =  \Tr [ \op{\varrho} \,\op{w}(q,p) ] \, , & \nonumber \\ 
  & &  \label{eq:HWWignerDef} \\
  & \op{\varrho}   = \displaystyle 
  \frac{1}{(2\pi)^{2}} \int_{\mathbb{R}^{2}}\op{w}(q,p) W(q,p) \, dq dp \, , & 
  \nonumber
\end{eqnarray}
where the (Hermitian) Wigner kernel $\op{w}$ (a particular instance of
a Stratonovitch-Weyl quantizer) is the double Fourier transform of the
displacement operator:
\begin{equation}
  \label{eq:HWkernelDef}
  \op{w} (q, p) = \frac{1}{(2\pi)^2} \int_{\mathbb{R}^2} 
  \exp[-i(p q^\prime -q p^\prime)] \op{D} (q^\prime, p^\prime) \, 
  dq^\prime dp^\prime \, .
\end{equation}
One can immediately check that the Wigner kernels are also a complete 
trace-orthonormal set. Furthermore, they transform properly under 
displacements
\begin{equation}
  \label{eq:HWKernelDisp}
  \op{w} (q, p) = \op{D} (q,p) \,\op{w} (0, 0) \,
  \op{D}^\dagger (q, p) \, ,
\end{equation}
where
\begin{equation}
  \label{eq:Parity}
  \op{w}(0,0)=\int_{\mathbb{R}^{2}} \op{D}(q, p) \, dq dp =  2 \op{P} \, ,
\end{equation} 
and $\op{P}$ is the parity operator.

The Wigner function in (\ref{eq:HWWignerDef}) fulfills all the basic
properties required for any good probabilistic description.  First,
due to the Hermiticity of $\op{w} (q,p)$, it is real for Hermitian
operators. Second, on integrating $W (q, p)$ over one variable, the 
probability distribution of the conjugate variable is reproduced
\begin{equation}
  \label{eq:HWProps2}
  \int_\mathbb{R}  W(q, p) \, dp = \langle q|\op{\varrho}| q \rangle \, ,
  \quad
  \int_\mathbb{R} W(q, p) \, dq = \langle p|\op{\varrho}| p \rangle \, .
\end{equation}
Third, $W(q, p)$ is covariant, which means that for the displaced 
state $\op{\varrho}^\prime = \op{D}(q_0, p_0) \,\op{\varrho} \, 
\op{D}^\dagger (q_0, p_0)$, one has
\begin{equation}
  \label{eq:HWProps3}
  W_{\op{\varrho}^\prime} (q, p) = W_{\op{\varrho}} (q-q_0, p-p_0) \, ,
\end{equation}
so that the Wigner function follows displacements rigidly without
changing its form, reflecting the fact that physics should not depend
on a certain choice of the origin.

Finally, the overlap of two density operators is proportional to
the integral of the associated Wigner functions:
\begin{equation}
  \label{eq:HWProps4}
  \Tr ( \op{\varrho}_1 \,\op{\varrho}_2 ) \propto
  \int_{\mathbb{R}^2} W_1(q, p)  W_2(q,p) \, dq dp \, .
\end{equation}
This property (often known as traciality) offers practical advantages, 
since it allows one to predict the statistics of any outcome, once the 
Wigner function of the measured state is known.

\section{Wigner function for  discrete systems}
\label{sec:Discrete}

Many quantum systems can be appropriately described in a
finite-dimensional Hilbert space. The previous standard approach can
be extended to these discrete systems, since they do have a dynamical
symmetry group.  However, in a continuous Wigner function for these
systems, there is a lot of information redundancy. The goal of this
section is to carry out a non-redundant discrete phase-space analysis
for this case.

Let us consider a system living in a Hilbert space $\mathcal{H}_{d}$,
of dimension $d$ (a qudit). It is useful to choose a computational
basis $ | n \rangle $ ($n = 0, \ldots , d-1$) in $\mathcal{H}_{d}$ and
introduce the basic operators~\cite{Schwinger:1960}
\begin{equation}
  \label{CC}
  \op{X} | n \rangle  =  |n + 1 \rangle \, ,
  \qquad 
  \op{Z} | n \rangle =  \omega(n) | n \rangle \, ,
\end{equation}
where addition and multiplication must be understood modulo $d$ and,
for simplicity, we use the notation
\begin{equation}
  \omega (m) \equiv \omega^{m} =  \exp (i 2\pi m/d) \, ,
\end{equation}
$\omega = \exp( i 2\pi/d)$ being a $d$th root of the unity. The
operators $\op{X}$ and $\op{Z}$ generate a group under multiplication
known as the generalized Pauli group~\cite{Nielsen:2000} and obey
\begin{equation}
  \label{eq:ZXwXZ}
  \op{Z} \op{X} = \omega \, \op{X} \op{Z} \, ,
\end{equation}
which is the finite-dimensional version of the Weyl form
(\ref{eq:Weyl}) of the commutation relations.

The monomials $\{ \hat{Z}^{k} \hat{X}^{l}\}$ ($k,l = 0, 1, \ldots, 
d-1$) form a basis in the space of all the operators acting in
$\mathcal{H}_{d}$~\cite{Klimov:2005}.  It seems then natural to 
introduce the unitary displacement operators
\begin{equation}
  \op{D}(k, l) = e^{i\phi (k,l)} \op{Z}^{k} \op{X}^{l} \, ,  
  \label{eq:desp}
\end{equation} 
where $\phi (k,l)$ is a phase. The unitarity condition imposes that
\begin{equation}
  \phi (k, l) + \phi (-k,-l) = - \frac{2\pi}{d} kl \, .  
\label{ph_condition}
\end{equation}
Different choices have been analyzed in the literature~\cite{Vourdas:2007}; 
one of special  relevance  is
\begin{equation}
  \phi ( k, l ) = \frac{2 \pi}{d} 2^{-1}  \, kl \, ,
  \label{phi1}
\end{equation}
where $2^{-1}$ is the multiplicative inverse of 2 in $\mathbb{Z}_{d}$
when d is prime and $2^{-1}=1/2$ for nonprime dimensions.

In this way, we have got a discrete phase space of the system as 
a $d\times d$ grid of points, in a such a way that the coordinate of 
each point $(k, l)$ define powers of  $Z$ (``position'') and $X$ 
(``momentum'')  and the whole phase space is isomorphic  to a discrete 
two-dimensional torus.

The following mapping from the Hilbert space into the discrete 
phase space [equivalent to (\ref{eq:Wigcan})] 
\begin{eqnarray}
& W(k, l)  =  \Tr [ \op{\varrho} \, \op{w}(k, l)] \, , & \nonumber \\
 & & \\
& \op{\varrho}  =  \displaystyle 
\frac{1}{d^{2}} \sum_{k,l} \op{w}(k, l) W(k, l) \, , & \nonumber 
\label{eq:Wigdis}
\end{eqnarray}
is established in terms of the following (Hermitian) Wigner kernel 
\begin{equation}
 \op{w}(k, l) = \frac{1}{d^{2}} 
 \sum_{m,n}\omega (kn-lm) \,\op{D}(m,n) \, , \label{eq:kernel}
\end{equation}
which is normalized, satisfies the overlap condition
\begin{equation}
  \Tr [ \op{w}(k, l) \op{w}(k^{\prime}, l^{\prime}) ] = d \, 
  \delta_{k, k^{\prime}} \, \delta_{l, l^{\prime}} \, ,
\end{equation}
and it is explicitly covariant:
\begin{equation}
  \op{w} (k, l) = \op{D} (k,l) \,\op{w}(0,0) \,\op{D}^\dagger (k, l) \, ,
  \label{eq:par}
\end{equation}
where
\begin{equation}
  \op{w}(0,0) = \frac{1}{d^{2}} \sum_{k,l} \op{D} (k, l) \, .  
  \label{eq:pari}
\end{equation}
It is interesting to note that the phase (\ref{phi1}) for prime
dimensions leads to $\op{w}(0,0) = \op{P}$, $\op{P}$ being the parity
operator.  In view of these properties, one can easily conclude that 
the corresponding Wigner function $W (k, l)$ fulfills properties 
fully analogous as those for the continuous harmonic oscillator.

\section{Wigner function for angle-angular momentum}
\label{sec:WigPhiL}

In this section, we consider the conjugate pair angle and angular
momentum. To avoid the difficulties linked with periodicity, the
simplest solution ~\cite{Louisell:1963,Mackey:1963,Carruthers:1968} 
is to adopt two angular coordinates, such as, e.g., cosine and sine, 
we shall denote by $\hat{C}$ and $\hat{S}$ to make no further assumptions
about the angle itself.  One can concisely condense all this
information using the complex exponential of the angle $\hat{E} =
\hat{C} + i \hat{S}$, which satisfies the commutation relation
\begin{equation}
  \label{ELE} 
  [ \hat{E},  \hat{L} ] = \hat{E} \, ,
\end{equation}
or, equivalently,
\begin{eqnarray}
  & [\op{C}, \op{L} ] = i \op{S} , 
  \qquad 
  [\op{S}, \op{L} ] = - i \op{C} \, , & \nonumber \\
  & & \\
  & [\op{C}, \op{S} ] = 0 \, . & \nonumber
\end{eqnarray}
In mathematical terms, this defines the Lie algebra of the
two-dimensional Euclidean group E(2). Note also, that from the
Baker-Campbell-Hausdorff formula, one gets
\begin{equation}
  \label{eq:ExpoCommute1}
  e^{-i\phi\op{L}} \op{E} = 
e^{i\phi} \, \op{E}  e^{- i\phi \op{L}} \, ,
\end{equation}
which is the unitary Weyl form of (\ref{ELE}).

The action of $\op{E}$ on the angular momentum basis is
\begin{equation}
  \label{E} 
  \hat{E} | \ell \rangle = | \ell - 1 \rangle \, ,
\end{equation}
and, since the integer $\ell$ runs from $- \infty$ to $+ \infty$,
$\op{E}$ is a unitary operator whose normalized eigenvectors
\begin{equation}
  \label{phi_states} 
  | \phi \rangle = \frac{1}{\sqrt{2 \pi}}
  \sum_{\ell \in \mathbb{Z}} e^{i \ell \phi} | \ell \rangle \, ,
\end{equation}
form a complete basis
\begin{equation}
\langle \phi | \phi^\prime \rangle = 
\sum_{\ell \in \mathbb{Z}} \delta (\phi - \phi^\prime -2 \ell \pi ) 
= \delta_{2\pi} (\phi - \phi^\prime ) \, ,
\label{eq:PhiNorm1}
\end{equation}
where $\delta_{2\pi}$ represents the periodic delta function (or
Dirac comb) of period $2 \pi$.

As anticipated in the Introduction, the phase space is now the 
semi-discrete cylinder $\mathbb{Z} \times  \mathcal{S}_1$. Following
the ideas of Sec.~\ref{sec:Discrete}, a displacement operator can be 
introduced as
\begin{equation}
  \label{eq:Displace1}
  \op{D} (\ell, \phi) = e^{i \alpha (\ell,\phi)} \, 
  \op {E}^{-\ell} e^{-i\phi \op{L}} \, ,
\end{equation}
where $ \alpha(\ell,\phi)$ is a phase to be specified. Note that
here there is no possibility to rewrite Eq.~(\ref{eq:Displace1}) as 
an entangled exponential, since the action of the operator to be
exponentiated would not be  well defined. The requirement of
unitarity imposes   now 
\begin{equation}
  \label{eq:AlphaCondition}
  \alpha(\ell, \phi) + \alpha(-\ell, - \phi) = \ell \phi \, . 
\end{equation}

As desired, the displacement operators form a complete
trace-orthonormal set: 
\begin{equation}
  \label{eq:DispOrtho}
  \Tr [ \op{D} ( \ell , \phi )  \op{D}^\dagger (\ell^\prime, \phi^\prime ) ] 
  =  2 \pi \, \delta_{\ell, \ell^\prime} \, \delta_{2\pi}(\phi -
  \phi^\prime) \, , 
\end{equation}
whose resemblance with relation~(\ref{eq:HWDispOrtho}) is evident.

We can introduce then the canonical mapping 
\begin{eqnarray}
&  W (\ell, \phi) = \Tr [ \op{\varrho} \,\op{w} ( \ell,\phi) ] \, , & 
\nonumber \\ 
& & \label{eq:WigFunDef1} \\
& \displaystyle
\op{\varrho} = \frac{1}{(2 \pi )^{2}} \, 
\suma{\ell} \intcirc \op{w}(\ell,\phi) W(\ell,\phi) \, d\phi \, , & 
\nonumber
\end{eqnarray}
where the Wigner kernel $\op{w}$ is defined, in close analogy to the
previous cases, as
\begin{equation}
  \label{eq:WigKerDef1}
  \op{w} (\ell, \phi) = 
  \frac{1}{(2\pi)^2} 
  \suma{\ell^\prime} \intcirc   
  \exp[-i ( \ell^\prime \phi - \ell \phi^\prime)] 
  \op{D} (\ell^\prime, \phi^\prime) \, d\phi^\prime \, .
\end{equation}

The set of Wigner kernels constitutes a complete orthogonal Hermitian 
operator basis. In addition, they are  explicitly covariant: 
\begin{equation}
  \label{eq:WigKerDisp}
  \op{w} (\ell,\phi) = \op{D}(\ell,\phi) \, 
  \op{w}(0,0) \,\op{D}^\dagger (\ell,\phi) \, ,
\end{equation}
with
\begin{equation}
\op{w} (0,0) = \frac{1}{(2\pi)^2} \suma{\ell} \intcirc  
\op{D} (\ell,\phi) \, d\phi \, ,
\end{equation}
although the interpretation of $\op{w} (0,0)$ as the parity on the
cylinder is problematic.

All these properties automatically guarantee that we have
indeed a well-behaved Wigner function for this canonical pair.

\section{Examples}
\label{sec:Examples}

To work out the explicit form of the Wigner function for a given 
state,  one first needs to specify the phase $\alpha(\ell,\phi)$ 
in  Eq.~(\ref{eq:AlphaCondition}).  For convenience, in this paper 
the choice 
\begin{equation}
  \label{eq:AlphaExplicit}
  \alpha (\ell,\phi) = - \ell \phi /2 
\end{equation}
shall be used, as it is linear in both arguments, and it appears 
to be the simplest function fulfilling the unitarity 
condition and the periodicity in $\phi$~\cite{Rigas:2008}.  

In this case, the Wigner kernel~(\ref{eq:WigKerDef1}) becomes
\begin{eqnarray}
  \label{eq:ExplicitKernel1}
  \op{w} (\ell,\phi) &  =  &  \displaystyle
  \frac{1}{(2\pi)^2} \suma{\ell^\prime,\ell^{\prime \prime}} 
  \intcirc  e^{i \ell^{\prime} \phi^{\prime}/2} \, 
  e^{-i\ell^{\prime \prime} \phi^{\prime}} \nonumber \\ 
  & \times & \displaystyle 
  e^{i(\ell \phi^{\prime} - \ell^{\prime} \phi)}
  |\ell^{\prime \prime} \rangle \langle \ell^{\prime \prime} -
  \ell^{\prime}| \,  
  d\phi^\prime \, . 
\end{eqnarray}
After some manipulations, we obtain 
\begin{eqnarray}
  \label{eq:ExplictKernel3}
 \op{w} (\ell,\phi) & = & \displaystyle
 \frac{1}{2\pi} \suma{\ell^\prime}  e^{-2 i \ell^\prime \phi}
 |\ell + \ell^\prime \rangle  \langle \ell - \ell^\prime | \nonumber  \\
& + & \displaystyle
\frac{1}{2\pi^2} \suma{\ell^{\prime},\ell^{\prime \prime}} 
\frac{(-1)^{\ell^{\prime \prime}}}{\ell^{\prime \prime}+1/2} 
e^{-(2 \ell^{\prime} + 1) i \phi} 
\nonumber \\
& \times & 
 |\ell + \ell^{\prime \prime}+\ell^{\prime}+1\rangle
\langle\ell +\ell^{\prime \prime}-\ell^{\prime}| \, ,
\end{eqnarray}
which coincides with the kernel derived by Plebanski and 
coworkers~\cite{Plebanski:2000} in the context of deformation 
quantization. 

Note that (\ref{eq:ExplictKernel3}) splits into ``even'' and  
``odd'' parts, depending on whether the matrix elements 
$\varrho_{\ell \ell^{\prime}} = \langle \ell| \op{\varrho} | 
\ell^{\prime}\rangle$ have $\ell \pm \ell^{\prime}$ even 
(first sum)  or odd (second sum).

For an angular momentum eigenstate $| \ell_0 \rangle$, one immediately
gets
\begin{equation}
  \label{eq:ExampleOAMstaleEll}
  W_{| \ell_0 \rangle} (\ell,\phi) = \frac{1}{2\pi} 
  \delta_{\ell, \ell_0} \, , 
\end{equation}
which is quite reasonable in this case: it is  flat in $\phi$ and 
the integral over the whole phase space gives the unity, reflecting 
the normalization of $|\ell_0\rangle$.

For an angle eigenstate $|\phi_0 \rangle$, one has 
\begin{equation} 
  \label{eq:ExamlpeE-StatePhi2} 
 W_{| \phi_0 \rangle} (\ell, \phi) = \frac{1}{2\pi} \,
 \delta_{2\pi}(\phi-\phi_0) \, . 
\end{equation} 
Now, the Wigner function is flat in the conjugate variable $\ell$, 
and thus, the integral over the whole phase space diverges, 
which is a consequence of the fact that the state $|\phi_0\rangle$
is not normalizable. 

The  coherent states  $| \ell_0, \phi_0 \rangle$ (parametrized by 
points on the cylinder) introduced in Ref.~\cite{Kowalski:1996} 
(see also Refs.~\cite{Gonzalez:1998,Kastrup:2006} for a detailed 
discussion of the properties of these relevant states) are 
characterized by
\begin{eqnarray}
  \langle \ell | \ell_0 , \phi_0 \rangle & =  &
  \displaystyle
  \frac{1}{\sqrt{\vartheta_3 \left ( 0  \big | \frac{1}{e} \right )}}
  e^{-i \ell \phi_0} \, e^{-(\ell - \ell_0)^2/2} \, , \nonumber  \\
 \label{eq:ExampleCoh1}
  & & \\
  \langle \phi| \ell_0 , \phi_0 \rangle & = &
  \displaystyle
  \frac{e^{i \ell_0 (\phi - \phi_0)}}
  {\sqrt{\vartheta_3 \left ( 0  \big | \frac{1}{e} \right )}}
  \vartheta_3\left(\frac{\phi-\phi_0}{2} \Big | \frac{1}{e^2} \right) ,
  \nonumber
\end{eqnarray}
where $\vartheta_3$ denotes the third Jacobi theta 
function~\cite{Mumford:1983}.

The Wigner function for the state $|\ell_0,\phi_0\rangle$ splits as
\begin{equation}
  \label{eq:ExampleWCoh}
  W_{|\ell_0,\phi_0 \rangle} (\ell,\phi) = 
  W^{(+)}_{|\ell_0,\phi_0\rangle} (\ell, \phi) +  
  W^{(-)}_{|\ell_0,\phi_0\rangle} (\ell, \phi) \, .
\end{equation}
The ``even'' part turns out to be
\begin{equation}
  W^{(+)}_{|\ell_0,\phi_0\rangle} (\ell,\phi) =  
  \frac{1}{2 \pi \vartheta_3 \left (0 \big |\frac{1}{e} \right )} 
  e^{-(\ell -\ell_0)^2} 
  \vartheta_3 \left(\phi-\phi_0 \Big|  \frac{1}{e} \right) \, .
\end{equation}
This seems a sensible result, since it is a discrete Gaussian in the
variable $\ell$, and for the continuous angle $\phi$ it is a Jacobi
theta function, which plays the role of the Gaussian for circular
statistics~\cite{Rehacek:2008}. However, the ``odd'' contribution 
spoils this simple picture:
\begin{eqnarray}
  \nonumber
  W^{(-)}_{|\ell_0,\phi_0\rangle} (\ell,\phi) &=&
  \frac{e^{i(\phi-\phi_0)-1/2}}{2\pi^2 \vartheta_3
    \left ( 0  \big | \frac{1}{e} \right )}
  \vartheta_3\left( \phi- \phi_0 +i/2 \Big|\frac{1}{e} \right) \nonumber \\
  & \times &
  \suma{\ell^\prime} (-1)^{\ell^\prime - \ell + \ell_0} 
  \frac{e^{-\ell^\prime {}^2 - \ell^\prime}}{\ell^\prime + \ell_0 - \ell + 1/2} \, .
  \label{eq:ExampleWoddCoh}  
\end{eqnarray}

\begin{figure}[t]
\includegraphics[width=0.80\columnwidth]{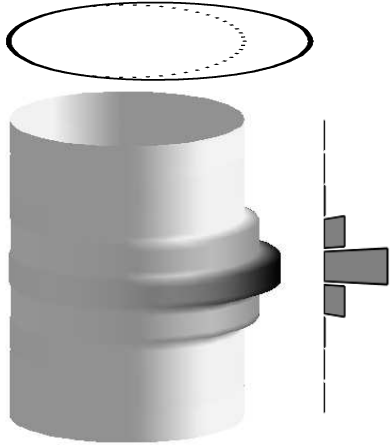}
\caption{Plot of the Wigner function for a coherent state with
$\ell_0 =0$ and $\phi_0 = 0$. The cylinder extends vertically
from $\ell = -4 $ to $\ell = +4$. The two
corresponding 
marginal distributions are shown.}
\label{fig:CoherentStandard}
\end{figure}

In Fig.~\ref{fig:CoherentStandard}, the Wigner function for the
coherent state $| \ell_0 = 0, \phi_0 = 0\rangle$ is plotted on the
discrete cylinder.  A pronounced peak at $\phi=0$ for $\ell=0$ and
slightly smaller ones for $\ell=\pm 1$ can be observed.  The associated
marginal distributions [obtained from Eq.~(\ref{eq:ExampleWCoh}) by 
integrating over $\phi$ or  by summing over $\ell$, respectively] 
are also plotted. They are strictly positive, as correspond to
true probability distributions.

\begin{figure}[b]
\includegraphics[width=0.90\columnwidth]{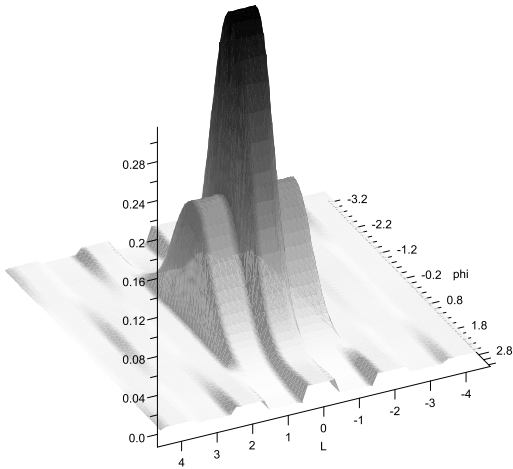}
\caption{Unwrapped plot of the Wigner function for a coherent state
  with 
$\ell_0 =0$ and $\phi_0 = 0$. The plane extends 
from $\ell = -4 $ to $\ell = +4$ and from $\phi=-\pi$ to
$\phi=\pi$.}
\label{fig:CoherentPico}
\end{figure}

For quantitative comparisons, however, sometimes it may be convenient 
to ``cut'' this cylindrical plot along a line $\phi$=constant and
unwrap it. This is shown in Fig.~\ref{fig:CoherentPico}. Here, the 
range of $\ell$ is from -4 to 4, while the angle is plotted between 
$-\pi$ to $\pi$.

A closer look at these figures reveals also a remarkable fact: for
values close to $\phi=\pm \pi$ and $\ell= \pm 1$, the Wigner function
takes negative values.  Actually, a numeric analysis suggests the
existence of negativities close to $\phi = \pm \pi$ for any odd value
of $\ell$.

\begin{figure}[t]
\includegraphics[width=0.80\columnwidth]{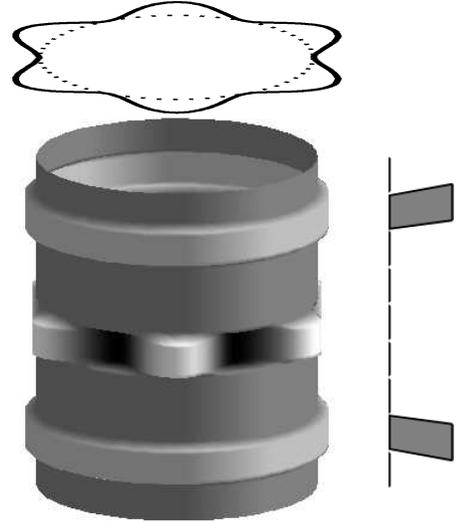}
\caption{Plot and marginal distributions
of the Wigner function for an even  superposition 
$|\ell_1 +_\theta\ell_2\rangle $ with $\ell_{1,2}=\pm 3$ 
for  $\ell = -4 $ to $\ell = +4$.}
\label{fig:EvenTornillo}
\end{figure}

\begin{figure}[b]
\includegraphics[width=0.80\columnwidth]{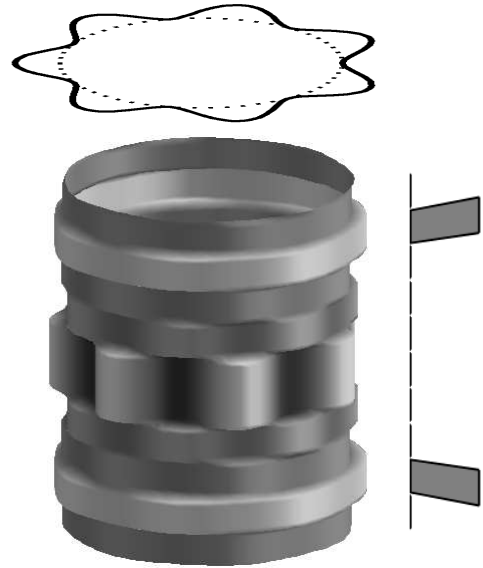}
\caption{Plot and marginal distributions
of the Wigner function for an even  superposition 
$|\ell_1 +_\theta\ell_2\rangle $ with $\ell_1= 4, \ell_2=-3$ 
for  $\ell = -4 $ to $\ell = +5$.}
\label{fig:OddTornillo}
\end{figure}

As our last example, we address the superposition
\begin{equation}
  \label{eq:SuperDef}
  |\Psi \rangle = \frac{1}{\sqrt{2}} 
  (|\ell_1\rangle   + e^{i\phi_0} | \ell_2\rangle )
\end{equation}
of two angular-momentum eigenstates with a relative phase 
$e^{i \phi_0}$. The analysis can be carried out for the superposition 
of any number of eigenstates, but (\ref{eq:SuperDef}) is enough to
display the relevant features.

The Wigner function splits again;  now the  ``even'' part reads as
\begin{eqnarray}
  \label{eq:ExampleSuperEven}
  W_{  | \Psi \rangle}^{(+)} (\ell, \phi) &  = &  
  \frac{1}{4\pi}  \{ \delta_{\ell, \ell_1} + \delta_{\ell, \ell_2} \nonumber \\
 & + & 
  2 \delta_{\ell_ 1+\ell_2, 2\ell} \, \cos[\phi_0 + (\ell_2 - \ell_1) \phi]
  \} \, .  
\end{eqnarray}
For the ``odd'' part, the diagonal contributions vanish, and one has 
\begin{eqnarray} 
W_{  |\Psi \rangle}^{(-)} (\ell, \phi) &  =  & \displaystyle 
\frac{1}{\pi^2} \cos[\phi_0 + (\ell_2 - \ell_1) \phi] \nonumber \\
& \times &   \displaystyle 
\frac{(-1)^{\ell+(\ell_1+\ell_2 -1)/2}}{\ell_1 +\ell_2 - 2\ell} 
\delta_{\ell_1 + \ell_2 = \mathrm{odd}} \, ,
\label{eq:ExamplesSuperOdd2}
\end{eqnarray}
where $\delta_{\ell_1 + \ell_2 = \mathrm{odd}}$  indicates that the sum 
is nonzero only when $\ell_1 + \ell_2$ is odd.

In consequence, when $| \ell_1 - \ell_2|$ is odd, the interference
term contains contributions for any $\ell$, damped as $1/\ell$.  
When $|\ell_1 - \ell_2|$ is an even number, the
contribution~(\ref{eq:ExamplesSuperOdd2}) vanishes and we have 
three contributions: two flat slices coming from the states 
$| \ell_1\rangle$ and $| \ell_2\rangle$ and an interference term 
located at $\ell = (\ell_1 + \ell_2)/2$.

These features are illustrated in Figs.~\ref{fig:EvenTornillo} and
\ref{fig:OddTornillo}. The state $|\Psi \rangle $ is plotted for 
$\ell_2 = -3$ and $\ell_1 = 3$ and (Fig.~\ref{fig:EvenTornillo}) 
and $\ell_2 = -3$ and $\ell_1 = 4$  (Fig.~\ref{fig:OddTornillo}). 
Changing the relative phase $\phi_0$ results in a global rotation
of the cylinder.  In can be observed in Fig.~\ref{fig:OddTornillo}
that the two rings at $\ell = -3$ and $\ell=4$ (as opposed to the rings 
at  $\ell = \pm 3$ in Fig.~\ref{fig:EvenTornillo}), are not flat in $\phi$,
but show a weak dependence on the angle due to the odd contributions added 
to the flat Kronecker deltas.

\section{Concluding remarks}

In summary, we have carried out a full program for a complete
phase-space description in terms a Wigner function for the canonical
pair angle-angular momentum. An experimental demonstration in terms of
optical beams is presently underway in our laboratory.

\acknowledgments

We acknowledge discussions with Hubert de Guise, Jose Gracia-Bondia,
Hans Kastrup, Jakub Rembielinski and Krzysztof Kowalski.  This work
was supported by the Czech Ministry of Education, Project
MSM6198959213, the Czech Grant Agency, Grant 202/06/0307, 
the Spanish Research Directorate, Grant FIS2005-06714, and the
Mexican  CONACYT, Grant 45705.


\begin{thebibliography}{41}
\expandafter\ifx\csname natexlab\endcsname\relax\def\natexlab#1{#1}\fi
\expandafter\ifx\csname bibnamefont\endcsname\relax
  \def\bibnamefont#1{#1}\fi
\expandafter\ifx\csname bibfnamefont\endcsname\relax
  \def\bibfnamefont#1{#1}\fi
\expandafter\ifx\csname citenamefont\endcsname\relax
  \def\citenamefont#1{#1}\fi
\expandafter\ifx\csname url\endcsname\relax
  \def\url#1{\texttt{#1}}\fi
\expandafter\ifx\csname urlprefix\endcsname\relax\def\urlprefix{URL }\fi
\providecommand{\bibinfo}[2]{#2}
\providecommand{\eprint}[2][]{\url{#2}}

\bibitem[{\citenamefont{Barut and R{\c{a}}czka}(1987)}]{Barut:1987}
\bibinfo{author}{\bibfnamefont{A.~O.} \bibnamefont{Barut}} \bibnamefont{and}
  \bibinfo{author}{\bibfnamefont{R.}~\bibnamefont{R{\c{a}}czka}},
  \emph{\bibinfo{title}{Theory of Group Representations and Applications}}
  (\bibinfo{publisher}{World Scientific}, \bibinfo{address}{Singapore},
  \bibinfo{year}{1987}).

\bibitem[{\citenamefont{Kostant}(1970)}]{Kostant:1970}
\bibinfo{author}{\bibfnamefont{B.}~\bibnamefont{Kostant}},
  \bibinfo{journal}{Lect. Notes Math.} \textbf{\bibinfo{volume}{170}},
  \bibinfo{pages}{87} (\bibinfo{year}{1970}).

\bibitem[{\citenamefont{Kirillov}(1976.)}]{Kirillov:1976}
\bibinfo{author}{\bibfnamefont{A.~A.} \bibnamefont{Kirillov}},
  \emph{\bibinfo{title}{Elements of the Theory of Representations}}
  (\bibinfo{publisher}{Springer-Verlag}, \bibinfo{address}{Berlin},
  \bibinfo{year}{1976.}).

\bibitem[{\citenamefont{Weyl}(1950)}]{Weyl:1928}
\bibinfo{author}{\bibfnamefont{H.}~\bibnamefont{Weyl}},
  \emph{\bibinfo{title}{Gruppentheorie und Quantenmechanik}}
  (\bibinfo{publisher}{Hirzel}, \bibinfo{address}{Leipzig},
  \bibinfo{year}{1950}).

\bibitem[{\citenamefont{Wigner}(1932)}]{Wigner:1932}
\bibinfo{author}{\bibfnamefont{E.~P.} \bibnamefont{Wigner}},
  \bibinfo{journal}{Phys. Rev.} \textbf{\bibinfo{volume}{40}},
  \bibinfo{pages}{749} (\bibinfo{year}{1932}).

\bibitem[{\citenamefont{Moyal}(1949)}]{Moyal:1949}
\bibinfo{author}{\bibfnamefont{J.~E.} \bibnamefont{Moyal}},
  \bibinfo{journal}{Proc. Camb. Phil. Soc.} \textbf{\bibinfo{volume}{45}},
  \bibinfo{pages}{99} (\bibinfo{year}{1949}).

\bibitem[{\citenamefont{Stratonovich}(1956)}]{Stratonovich:1956}
\bibinfo{author}{\bibfnamefont{R.~L.} \bibnamefont{Stratonovich}},
  \bibinfo{journal}{JETP} \textbf{\bibinfo{volume}{31}}, \bibinfo{pages}{1012}
  (\bibinfo{year}{1956}).

\bibitem[{\citenamefont{Agarwal and Wolf}(1970)}]{Agarwal:1970}
\bibinfo{author}{\bibfnamefont{G.~S.} \bibnamefont{Agarwal}} \bibnamefont{and}
  \bibinfo{author}{\bibfnamefont{E.}~\bibnamefont{Wolf}},
  \bibinfo{journal}{Phys. Rev. D} \textbf{\bibinfo{volume}{2}},
  \bibinfo{pages}{2161 } (\bibinfo{year}{1970}).

\bibitem[{\citenamefont{Berezin}(1975)}]{Berezin:1975}
\bibinfo{author}{\bibfnamefont{F.~A.} \bibnamefont{Berezin}},
  \bibinfo{journal}{Commun. Math. Phys.} \textbf{\bibinfo{volume}{40}},
  \bibinfo{pages}{153} (\bibinfo{year}{1975}).

\bibitem[{\citenamefont{Agarwal}(1981)}]{Agarwal:1981}
\bibinfo{author}{\bibfnamefont{G.~S.} \bibnamefont{Agarwal}},
  \bibinfo{journal}{Phys. Rev. A} \textbf{\bibinfo{volume}{24}},
  \bibinfo{pages}{2889} (\bibinfo{year}{1981}).

\bibitem[{\citenamefont{Bertrand and Bertrand}(1987)}]{Bertrand:1987}
\bibinfo{author}{\bibfnamefont{J.}~\bibnamefont{Bertrand}} \bibnamefont{and}
  \bibinfo{author}{\bibfnamefont{P.}~\bibnamefont{Bertrand}},
  \bibinfo{journal}{Found. Phys.} \textbf{\bibinfo{volume}{17}},
  \bibinfo{pages}{397} (\bibinfo{year}{1987}).

\bibitem[{\citenamefont{Varilly and Gracia-Bond{\'{\i}}a}(1989)}]{Varilly:1989}
\bibinfo{author}{\bibfnamefont{J.~C.} \bibnamefont{Varilly}} \bibnamefont{and}
  \bibinfo{author}{\bibfnamefont{J.~M.} \bibnamefont{Gracia-Bond{\'{\i}}a}},
  \bibinfo{journal}{Ann. Phys. (NY)nn. Phy} \textbf{\bibinfo{volume}{190}},
  \bibinfo{pages}{107} (\bibinfo{year}{1989}).

\bibitem[{\citenamefont{Atakishiyev et~al.}(1998)\citenamefont{Atakishiyev,
  Chumakov, and Wolf}}]{Atakishiyev:1998}
\bibinfo{author}{\bibfnamefont{N.~M.} \bibnamefont{Atakishiyev}},
  \bibinfo{author}{\bibfnamefont{S.~M.} \bibnamefont{Chumakov}},
  \bibnamefont{and} \bibinfo{author}{\bibfnamefont{K.~B.} \bibnamefont{Wolf}},
  \bibinfo{journal}{J. Math. Phys.} \textbf{\bibinfo{volume}{39}},
  \bibinfo{pages}{6247} (\bibinfo{year}{1998}).

\bibitem[{\citenamefont{Brif and Mann}(1998)}]{Brif:1998}
\bibinfo{author}{\bibfnamefont{C.}~\bibnamefont{Brif}} \bibnamefont{and}
  \bibinfo{author}{\bibfnamefont{A.}~\bibnamefont{Mann}}, \bibinfo{journal}{J.
  Phys. A} \textbf{\bibinfo{volume}{31}}, \bibinfo{pages}{L9}
  (\bibinfo{year}{1998}).

\bibitem[{\citenamefont{Benedict and Czirj{\'a}k}(1999)}]{Benedict:1999}
\bibinfo{author}{\bibfnamefont{M.~G.} \bibnamefont{Benedict}} \bibnamefont{and}
  \bibinfo{author}{\bibfnamefont{A.}~\bibnamefont{Czirj{\'a}k}},
  \bibinfo{journal}{Phys. Rev. A} \textbf{\bibinfo{volume}{60}},
  \bibinfo{pages}{4034} (\bibinfo{year}{1999}).

\bibitem[{\citenamefont{Balazs and Jennings}(1984)}]{Balazs:1984}
\bibinfo{author}{\bibfnamefont{N.~L.} \bibnamefont{Balazs}} \bibnamefont{and}
  \bibinfo{author}{\bibfnamefont{B.~K.} \bibnamefont{Jennings}},
  \bibinfo{journal}{Phys. Rep.} \textbf{\bibinfo{volume}{104}},
  \bibinfo{pages}{347} (\bibinfo{year}{1984}).

\bibitem[{\citenamefont{Hillery et~al.}(1984)\citenamefont{Hillery, Connell,
  Scully, and Wigner}}]{Hillery:1984}
\bibinfo{author}{\bibfnamefont{M.}~\bibnamefont{Hillery}},
  \bibinfo{author}{\bibfnamefont{R.~F.~O.} \bibnamefont{Connell}},
  \bibinfo{author}{\bibfnamefont{M.~O.} \bibnamefont{Scully}},
  \bibnamefont{and} \bibinfo{author}{\bibfnamefont{E.~P.}
  \bibnamefont{Wigner}}, \bibinfo{journal}{Phys. Rep.}
  \textbf{\bibinfo{volume}{106}}, \bibinfo{pages}{121} (\bibinfo{year}{1984}).

\bibitem[{\citenamefont{Lee}(1995)}]{Lee:1995}
\bibinfo{author}{\bibfnamefont{H.-W.} \bibnamefont{Lee}},
  \bibinfo{journal}{Phys. Rep.} \textbf{\bibinfo{volume}{259}},
  \bibinfo{pages}{147} (\bibinfo{year}{1995}).

\bibitem[{\citenamefont{Jr.}(1996)}]{Schroek:1996}
\bibinfo{author}{\bibfnamefont{F.~E.~S.} \bibnamefont{Jr.}},
  \emph{\bibinfo{title}{Quantum Mechanics on Phase Space}}
  (\bibinfo{publisher}{Kluwer}, \bibinfo{address}{Dordrecht},
  \bibinfo{year}{1996}).

\bibitem[{\citenamefont{Klimov et~al.}(2008)\citenamefont{Klimov, Bj{\"{o}}rk,
  and S{\'{a}}nchez-Soto}}]{Klimov:2008}
\bibinfo{author}{\bibfnamefont{A.~B.} \bibnamefont{Klimov}},
  \bibinfo{author}{\bibfnamefont{G.}~\bibnamefont{Bj{\"{o}}rk}},
  \bibnamefont{and} \bibinfo{author}{\bibfnamefont{L.~L.}
  \bibnamefont{S{\'{a}}nchez-Soto}}, \bibinfo{journal}{Prog. Opt.}
  \textbf{\bibinfo{volume}{51}}, \bibinfo{pages}{469} (\bibinfo{year}{2008}).

\bibitem[{\citenamefont{Rigas et~al.}(2008)\citenamefont{Rigas,
  S{\'{a}}nchez-Soto, Klimov, \v{R}eh\'a\v{c}ek, and Hradil}}]{Rigas:2008}
\bibinfo{author}{\bibfnamefont{I.}~\bibnamefont{Rigas}},
  \bibinfo{author}{\bibfnamefont{L.~L.} \bibnamefont{S{\'{a}}nchez-Soto}},
  \bibinfo{author}{\bibfnamefont{A.~B.} \bibnamefont{Klimov}},
  \bibinfo{author}{\bibfnamefont{J.}~\bibnamefont{\v{R}eh\'a\v{c}ek}},
  \bibnamefont{and} \bibinfo{author}{\bibfnamefont{Z.}~\bibnamefont{Hradil}},
  \bibinfo{journal}{Phys. Rev. A} \textbf{\bibinfo{volume}{78}}
  (\bibinfo{year}{2008}).

\bibitem[{\citenamefont{Molina-Terriza
  et~al.}(2007)\citenamefont{Molina-Terriza, Torres, and Torner}}]{Molina:2007}
\bibinfo{author}{\bibfnamefont{G.}~\bibnamefont{Molina-Terriza}},
  \bibinfo{author}{\bibfnamefont{J.~P.} \bibnamefont{Torres}},
  \bibnamefont{and} \bibinfo{author}{\bibfnamefont{L.}~\bibnamefont{Torner}},
  \bibinfo{journal}{Nat. Phys.} \textbf{\bibinfo{volume}{3}},
  \bibinfo{pages}{305} (\bibinfo{year}{2007}).

\bibitem[{\citenamefont{Franke-Arnold et~al.}(2008)\citenamefont{Franke-Arnold,
  Allen, and Padgett}}]{Franke:2008}
\bibinfo{author}{\bibfnamefont{S.}~\bibnamefont{Franke-Arnold}},
  \bibinfo{author}{\bibfnamefont{L.}~\bibnamefont{Allen}}, \bibnamefont{and}
  \bibinfo{author}{\bibfnamefont{M.}~\bibnamefont{Padgett}},
  \bibinfo{journal}{Laser Photon. Rev.} \textbf{\bibinfo{volume}{2}},
  \bibinfo{pages}{299} (\bibinfo{year}{2008}).

\bibitem[{\citenamefont{Vaziri et~al.}(2002)\citenamefont{Vaziri, Weihs, and
  Zeilinger}}]{Vaziri:2002}
\bibinfo{author}{\bibfnamefont{A.}~\bibnamefont{Vaziri}},
  \bibinfo{author}{\bibfnamefont{G.}~\bibnamefont{Weihs}}, \bibnamefont{and}
  \bibinfo{author}{\bibfnamefont{A.}~\bibnamefont{Zeilinger}},
  \bibinfo{journal}{J. Opt. B} \textbf{\bibinfo{volume}{4}},
  \bibinfo{pages}{S47} (\bibinfo{year}{2002}).

\bibitem[{\citenamefont{Mukunda}(1979)}]{Mukunda:1979}
\bibinfo{author}{\bibfnamefont{N.}~\bibnamefont{Mukunda}},
  \bibinfo{journal}{Am. J. Phys.} \textbf{\bibinfo{volume}{47}},
  \bibinfo{pages}{182} (\bibinfo{year}{1979}).

\bibitem[{\citenamefont{Bizarro}(1994)}]{Bizarro:1994}
\bibinfo{author}{\bibfnamefont{J.~P.} \bibnamefont{Bizarro}},
  \bibinfo{journal}{Phys. Rev. A} \textbf{\bibinfo{volume}{49}},
  \bibinfo{pages}{3255} (\bibinfo{year}{1994}).

\bibitem[{\citenamefont{Mukunda et~al.}(2005)\citenamefont{Mukunda, Marmo,
  Zampini, Chaturvedi, and Simon}}]{Mukunda:2005}
\bibinfo{author}{\bibfnamefont{N.}~\bibnamefont{Mukunda}},
  \bibinfo{author}{\bibfnamefont{G.}~\bibnamefont{Marmo}},
  \bibinfo{author}{\bibfnamefont{A.}~\bibnamefont{Zampini}},
  \bibinfo{author}{\bibfnamefont{S.}~\bibnamefont{Chaturvedi}},
  \bibnamefont{and} \bibinfo{author}{\bibfnamefont{R.}~\bibnamefont{Simon}},
  \bibinfo{journal}{J. Math. Phys.} \textbf{\bibinfo{volume}{46}},
  \bibinfo{pages}{012106} (\bibinfo{year}{2005}).

\bibitem[{\citenamefont{Nieto et~al.}(1998)\citenamefont{Nieto, Atakishiyev,
  Chumakov, and Wolf}}]{Nieto:1998}
\bibinfo{author}{\bibfnamefont{L.~M.} \bibnamefont{Nieto}},
  \bibinfo{author}{\bibfnamefont{N.~A.} \bibnamefont{Atakishiyev}},
  \bibinfo{author}{\bibfnamefont{S.~M.} \bibnamefont{Chumakov}},
  \bibnamefont{and} \bibinfo{author}{\bibfnamefont{K.~B.} \bibnamefont{Wolf}},
  \bibinfo{journal}{J. Phys. A} \textbf{\bibinfo{volume}{31}},
  \bibinfo{pages}{3875} (\bibinfo{year}{1998}).

\bibitem[{\citenamefont{Pleba{\'{n}}ski
  et~al.}(2000)\citenamefont{Pleba{\'{n}}ski, Prazanowski, Tosiek, and
  Turrubiates}}]{Plebanski:2000}
\bibinfo{author}{\bibfnamefont{J.~F.} \bibnamefont{Pleba{\'{n}}ski}},
  \bibinfo{author}{\bibfnamefont{M.}~\bibnamefont{Prazanowski}},
  \bibinfo{author}{\bibfnamefont{J.}~\bibnamefont{Tosiek}}, \bibnamefont{and}
  \bibinfo{author}{\bibfnamefont{F.~K.} \bibnamefont{Turrubiates}},
  \bibinfo{journal}{Acta Phys. Pol. B} \textbf{\bibinfo{volume}{31}},
  \bibinfo{pages}{561} (\bibinfo{year}{2000}).

\bibitem[{\citenamefont{Schwinger}(1960)}]{Schwinger:1960}
\bibinfo{author}{\bibfnamefont{J.}~\bibnamefont{Schwinger}},
  \bibinfo{journal}{Proc. Natl. Acad. Sci. USA} \textbf{\bibinfo{volume}{46}},
  \bibinfo{pages}{570} (\bibinfo{year}{1960}).

\bibitem[{\citenamefont{Nielsen and Chuang}(2000)}]{Nielsen:2000}
\bibinfo{author}{\bibfnamefont{M.~A.} \bibnamefont{Nielsen}} \bibnamefont{and}
  \bibinfo{author}{\bibfnamefont{I.~L.} \bibnamefont{Chuang}},
  \emph{\bibinfo{title}{Quantum Computation and Quantum Information}}
  (\bibinfo{publisher}{Cambridge University Press},
  \bibinfo{address}{Cambridge}, \bibinfo{year}{2000}).

\bibitem[{\citenamefont{Klimov et~al.}(2005)\citenamefont{Klimov,
  S{{\'a}}nchez-Soto, and de~Guise}}]{Klimov:2005}
\bibinfo{author}{\bibfnamefont{A.~B.} \bibnamefont{Klimov}},
  \bibinfo{author}{\bibfnamefont{L.~L.} \bibnamefont{S{{\'a}}nchez-Soto}},
  \bibnamefont{and} \bibinfo{author}{\bibfnamefont{H.}~\bibnamefont{de~Guise}},
  \bibinfo{journal}{J. Phys. A} \textbf{\bibinfo{volume}{38}},
  \bibinfo{pages}{2747} (\bibinfo{year}{2005}).

\bibitem[{\citenamefont{Vourdas}(2007)}]{Vourdas:2007}
\bibinfo{author}{\bibfnamefont{A.}~\bibnamefont{Vourdas}}, \bibinfo{journal}{J.
  Phys. A} \textbf{\bibinfo{volume}{40}}, \bibinfo{pages}{R285}
  (\bibinfo{year}{2007}).

\bibitem[{\citenamefont{Louisell}(1963)}]{Louisell:1963}
\bibinfo{author}{\bibfnamefont{W.~H.} \bibnamefont{Louisell}},
  \bibinfo{journal}{Phys. Lett.} \textbf{\bibinfo{volume}{7}},
  \bibinfo{pages}{60} (\bibinfo{year}{1963}).

\bibitem[{\citenamefont{Mackey}(1963)}]{Mackey:1963}
\bibinfo{author}{\bibfnamefont{G.~W.} \bibnamefont{Mackey}},
  \emph{\bibinfo{title}{Mathematical Foundations of Quantum Mechanics}}
  (\bibinfo{publisher}{Benjamin}, \bibinfo{address}{New York},
  \bibinfo{year}{1963}).

\bibitem[{\citenamefont{Carruthers and Nieto}(1968)}]{Carruthers:1968}
\bibinfo{author}{\bibfnamefont{P.}~\bibnamefont{Carruthers}} \bibnamefont{and}
  \bibinfo{author}{\bibfnamefont{M.~M.} \bibnamefont{Nieto}},
  \bibinfo{journal}{Rev. Mod. Phys} \textbf{\bibinfo{volume}{40}},
  \bibinfo{pages}{411} (\bibinfo{year}{1968}).

\bibitem[{\citenamefont{Kowalski et~al.}(1996)\citenamefont{Kowalski,
  Rembieli{\'n}ski, and Papaloucas}}]{Kowalski:1996}
\bibinfo{author}{\bibfnamefont{K.}~\bibnamefont{Kowalski}},
  \bibinfo{author}{\bibfnamefont{J.}~\bibnamefont{Rembieli{\'n}ski}},
  \bibnamefont{and} \bibinfo{author}{\bibfnamefont{L.~C.}
  \bibnamefont{Papaloucas}}, \bibinfo{journal}{J. Phys. A}
  \textbf{\bibinfo{volume}{29}}, \bibinfo{pages}{4149} (\bibinfo{year}{1996}).

\bibitem[{\citenamefont{Gonz{\'a}lez and del Olmo}(1998)}]{Gonzalez:1998}
\bibinfo{author}{\bibfnamefont{J.~A.} \bibnamefont{Gonz{\'a}lez}}
  \bibnamefont{and} \bibinfo{author}{\bibfnamefont{M.~A.} \bibnamefont{del
  Olmo}}, \bibinfo{journal}{J. Phys. A} \textbf{\bibinfo{volume}{31}}
  (\bibinfo{year}{1998}).

\bibitem[{\citenamefont{Kastrup}(2006)}]{Kastrup:2006}
\bibinfo{author}{\bibfnamefont{H.~A.} \bibnamefont{Kastrup}},
  \bibinfo{journal}{Phys. Rev. A} \textbf{\bibinfo{volume}{73}},
  \bibinfo{pages}{052104} (\bibinfo{year}{2006}).

\bibitem[{\citenamefont{Mumford}(1983)}]{Mumford:1983}
\bibinfo{author}{\bibfnamefont{D.}~\bibnamefont{Mumford}},
  \emph{\bibinfo{title}{Tata Lectures on Theta I}}
  (\bibinfo{publisher}{Birkhauser}, \bibinfo{address}{Boston},
  \bibinfo{year}{1983}).

\bibitem[{\citenamefont{\v{R}eh\'a\v{c}ek
  et~al.}(2008)\citenamefont{\v{R}eh\'a\v{c}ek, Bouchal, {\v{C}}elechovsk\'{y},
  Hradil, and S{\'{a}}nchez-Soto}}]{Rehacek:2008}
\bibinfo{author}{\bibfnamefont{J.}~\bibnamefont{\v{R}eh\'a\v{c}ek}},
  \bibinfo{author}{\bibfnamefont{Z.}~\bibnamefont{Bouchal}},
  \bibinfo{author}{\bibfnamefont{R.}~\bibnamefont{{\v{C}}elechovsk\'{y}}},
  \bibinfo{author}{\bibfnamefont{Z.}~\bibnamefont{Hradil}}, \bibnamefont{and}
  \bibinfo{author}{\bibfnamefont{L.~L.} \bibnamefont{S{\'{a}}nchez-Soto}},
  \bibinfo{journal}{Phys. Rev. A} \textbf{\bibinfo{volume}{77}},
  \bibinfo{pages}{032110} (\bibinfo{year}{2008}).

\end{thebibliography}

\end{document}